\documentclass[12pt]{article}
\usepackage{graphicx}
\usepackage{amsmath}
\usepackage{cite}

\begin{document}
\begin{center}
  \Large{Approximate evolution for a system composed by two coupled
    Jaynes-Cummings Hamiltonians} \\
  \vspace{.52cm}
\large{I. Ramos-Prieto$^{1,2}$, A. Paredes$^{1}$,
  J. R\'ecamier$^{1}$ and  H. Moya-Cessa$^{2}$} \\
$^{1}$ Instituto de Ciencias F\'{\i}sicas, Universidad
  Nacional Aut\'onoma de M\'exico, Apdo. Postal 48-3, Cuernavaca,
  Morelos 62251, M\'exico. \\
  $^{2}$ Instituto Nacional de
  Astrof\'{\i}sica, \'Optica y Electr\'onica, Calle Luis Enrique Erro
  No. 1, Santa Mar\'{\i}a Tonanzintla, Puebla, 72840, M\'exico. \\
\end{center}

\begin{abstract}
  In this work we construct an approximate time evolution operator for
  a system composed by two coupled Jaynes-Cummings Hamiltonians. We
  express the full time evolution operator as a product of
  exponentials and we analyze the validity of our approximations
  contrasting our analytical results with those obtained by purely
  numerical methods.
\end{abstract}
\section{Introduction}
The interaction between quantized light and matter has attracted a
lot of interest over the years
\cite{Jaynes,Shore,Haroche1,Braak,Chen}. Probably the simplest
model to study such interaction is by means of the Jaynes-Cummings
(JC) model \cite{Jaynes,Shore} which, via the so-called rotating
wave approximation has an exact solution, its solutions have a
very good agreement with those of the more general Rabi model
\cite{Braak} that takes into account counter-rotating terms. The
JC model has allowed the generation of several non-classical
states, among them, superpositions of coherent states
\cite{Glauber}, also called Schr\"odinger cats
\cite{Kurizki,Schleich}, number states \cite{Dantsker} and
squeezed states \cite{Kuklinski}. Because its algebraic properties
are similar to the ones used in ion-laser interactions,
generalizations to nonlinear Jaynes-Cummings models have been
proposed \cite{Vogel1,Vogel2,cordero1,cordero2,santos1}. Moreover,
by passing two-level atoms through the cavity, micromasers have
been studied \cite{Fili,Power}. In such systems, each atom, after
interacting, i.e., entangling with the quantized field, when
exiting the cavity they carry the information with them,
information that may be lost (if the atoms are not measured) or
that may be extracted by means of conditional measurements
\cite{Kurizki,Dantsker} in order to produce several non-classical
fields.

In 1970, Moore studied the quantum theory of the electromagnetic field in a variable length one-dimensional cavity \cite{moore}; an important result was the prediction that real photons can be created from the vacuum due to the effect moving mirrors have on the zero-point energy of the field. This effect is now known as the Dynamical Casimir Effect (DCE) and is considered to be a direct proof of the existence of quantum vacuum fluctuations of the electromagnetic field \cite{nation}. It has been stated that the DCE provides a means to generate quantum correlations, for instance, in Ref~\cite{dodonov} the authors considered how the DCE dynamics is affected by the presence of two two-level atoms interacting with a single resonance cavity field mode in the absence of damping and they found that it is possible to generate entangled states. Cavity quantum electrodynamics provides a promising setting for the preparation of distributed entanglement due to the strong coupling between atoms and cavity and its good insulation against environment \cite{yan,doherty}.
The experimental realization of the DCE was achieved through the architecture of superconducting quantum circuits, where the effective length of the resonator is rapidly modulated \cite{hassel,nori}. 
Applications of the dynamical Casimir effect to create highly entangled states using quantum circuits was also proposed in reference \cite{solano}.

In this manuscript we study a generalization of the single
Jaynes-Cummings model. We consider  two cavities coupled  by the overlap of evanescent cavity fields, each with a two-level atom
inside, in the absence of damping, and such that photons may hop from one cavity to the other
\cite{Hanoura}. This is done by obtaining an effective Hamiltonian by using James method for treating dispersive regimes \cite{James}. Those systems are of interest as they may be used to transfer quantum information and therefore, being the smallest building block of a possible chain, it is of interest to investigate them in detail.  Here, we will show how to obtain a time evolution operator that can be written in a product form. In order to do that we will make some approximations at the level of the interaction Hamiltonian and use the Wei-Norman theorem to obtain the evolution operator that may be applied  to specific initial conditions such that we obtain approximate solutions. 

\section{Theory}
In Ref.~\cite{solano} the authors investigate how to generate
multipartite entangled states of two-level systems by means of
varying boundary conditions in the framework of superconducting
circuits. The model consists of two cavities coupled to
independent single qubits. These cavities share a partially
reflecting and transparent mirror yielding the last interaction
term of the Hamiltonian
\begin{equation}\label{eq:solano}
\hat H = \hbar \sum_{l=1}^{2} \left[ \omega_l \hat a^{\dagger}_l
\hat a_l +\frac{\Omega_l}{2} \hat\sigma_l^{(z)} + g_l (\hat
\sigma_{l}^{(+)} \hat a_l +\hat \sigma_{l}^{(-)} \hat
a_l^{\dagger}) \right]
  +\hbar \alpha(t)(\hat a_1^{\dagger}+\hat a_1)(\hat a_2^{\dagger}+\hat a_2)
\end{equation}
When the effective cavity length is oscillating with small
amplitude the cavity-cavity coupling parameter is a time dependent
function $\alpha(t) = \alpha_0 \cos(\omega_d t)$. In
Ref.\cite{solano} the authors take $\omega_d = \omega_1+\omega_2$
and $\alpha_0/\omega_i\ll 1$ and the cavity-cavity interaction can
be approximated as
\[ \hbar \alpha(t)(\hat a_1^{\dagger}+\hat a_1)(\hat
a_2^{\dagger}+\hat a_2) \simeq \hbar \frac{\alpha_0}{2}\left(\hat
a_1^{\dagger}\hat a_2^{\dagger}+\hat a_1 \hat a_2\right). \]

Here we are interested in mode mixing, then, the driving frequency
is chosen as $\omega_d = \omega_1-\omega_2$ and one gets an
approximate cavity-cavity interaction of the form:
\[ \hbar \alpha(t)(\hat a_1^{\dagger}+\hat a_1)(\hat
a_2^{\dagger}+\hat a_2) \simeq \hbar \frac{\alpha_0}{2}\left(\hat
a_1^{\dagger}\hat a_2+\hat a_1 \hat a_2^{\dagger}\right). \]

The system's Hamiltonian is:
\begin{eqnarray} \label{eq:hamiltonian}
\frac{\hat H}{\hbar} & = & \omega_1 \hat n_1 +
\frac{\Omega_1}{2}\hat \sigma_1^{(z)}  +  \omega_2 \hat n_2 +
\frac{\Omega_1}{2}\hat \sigma_2^{(z)} +  g_1(\hat a_1 \hat
\sigma_1^{(+)} + \hat a_1^{\dagger}\hat \sigma_{1}^{(-)}) \nonumber
\\ & + & g_2(\hat a_2\hat \sigma_{2}^{(+)} + \hat
a_2^{\dagger}\hat \sigma_{2}^{(-)}) + \lambda (\hat a_1\hat
a_2^{\dagger}+\hat a_1^{\dagger}\hat a_2) \equiv \frac{\hat
H_0}{\hbar} + \frac{\hat V}{\hbar} ,
\end{eqnarray}
where we take as the unperturbed Hamiltonian
\[ \hat H_0 = \hbar\left(\omega_1 \hat n_1 +
\frac{\Omega_1}{2}\hat \sigma_1^{(z)}  +  \omega_2 \hat n_2 +
\frac{\Omega_2}{2}\hat \sigma_2^{(z)} \right), \] whose time
evolution operator is
\[ \hat U_0 = e^{-i\omega_1 t \hat n_1} e^{-i\frac{\Omega_1}{2}t \hat
\sigma_{1}^{(z)}} e^{-i\omega_2 t \hat n_2}
e^{-i\frac{\Omega_2}{2}t
  \hat \sigma_{2}^{(z)}}, \]
and the interaction picture Hamiltonian is obtained from:
\[ \hat H_I = \hat U_0^{\dagger}\hat V \hat U_0. \]
By applying the transformation we get the Hamiltonian
\begin{eqnarray}
  \hat H_I & = & \hbar g_1 (\hat \sigma_{1}^{(+)}\hat a_1
e^{-i(\omega_1-\Omega_1)t}+\hat \sigma_{1}^{(-)} \hat
a_1^{\dagger}e^{i(\omega_1-\Omega_1)t})\nonumber \\ &+& \hbar g_2
(\hat \sigma_{2}^{(+)}\hat a_2 e^{-i(\omega_2-\Omega_2)t}+\hat
\sigma_{2}^{(-)} \hat a_2^{\dagger}e^{i(\omega_2-\Omega_2)t})
\nonumber \\ &+& \hbar\lambda \left(\hat a_1^{\dagger} \hat a_2
e^{i(\omega_1-\omega_2) t} + \hat a_1 \hat a_2^{\dagger} e^{-i(\omega_1-\omega_2)t}\right)
\end{eqnarray}

\begin{equation}
  \hat H_I = \hat V_1 + \hat V_2,
\end{equation}
with
\begin{equation}
  \hat V_1 = \hbar \lambda (\hat a_1 \hat
 a_2^{\dagger}e^{-i(\omega_1-\omega_2)t} + \hat a_1^{\dagger}\hat a_2
 e^{i(\omega_1-\omega_2)t}),
\end{equation}
and
\begin{eqnarray}
  \hat V_2 &=& \hbar g_1 (\hat \sigma_{1}^{(+)}\hat a_1
 e^{-i(\omega_1-\Omega_1)t}+\hat \sigma_{1}^{(-)} \hat
 a_1^{\dagger}e^{i(\omega_1-\Omega_1)t})\nonumber \\ &+& \hbar g_2 (\hat
 \sigma_{2}^{(+)}\hat a_2 e^{-i(\omega_2-\Omega_2)t}+\hat
 \sigma_{2}^{(-)} \hat a_2^{\dagger}e^{i(\omega_2-\Omega_2)t}),
\end{eqnarray}
 such that $\hat U =\hat U_0\hat U_I = \hat U_0 \hat U_I^{(1)} \hat U_I^{(2)} $.

By defining the operators:
\[ \hat J_{+} = \hat a_1 \hat a_2^{\dagger}, \ \ \ \hat J_{-} = \hat
a_{1}^{\dagger}\hat a_2, \ \ \ \hat J_z = \hat a_1^{\dagger}\hat
a_1 -\hat a_2^{\dagger}\hat a_2, \] we can write
\[ \hat V_1 = \hbar \lambda \left(
e^{-i(\omega_1-\omega_2)t}\hat J_{+} +
e^{i(\omega_1-\omega_2)t}\hat J_{-} \right). \] The operators
$\hat J_{+}$, $\hat J_{-}$ have the commutation relations
\[ [\hat J_{-},\hat J_{+}] = \hat J_z, \ \ \ [\hat J_{-},\hat J_{z}]
=-2\hat J_{-}, \ \ \ [\hat J_{+},\hat J_z] = 2\hat J_{+}, \] with
$\hat{J}_z= \hat n_1 -\hat n_2$ and $\hat n_i=\hat a_i^{\dagger}\hat
a_i$. Since the interaction $\hat V_1$ is closed under commutation
we can apply the Wei-Norman theorem \cite{wei-norman} and write
the time evolution operator  as:
\begin{equation}\label{eq:opev1}
  \hat U_I^{(1)} = e^{\gamma_1 \hat J_{+}} e^{\gamma_2 \hat J_{-}}
  e^{\gamma_3 \hat J_z},
  \end{equation}
with:

\begin{eqnarray*}
  \dot \gamma_1 = -i \lambda
   \left(e^{-i(\omega_1-\omega_2)t}-\gamma_1^2
 e^{i(\omega_1-\omega_2)t}\right)  \\
 \dot \gamma_2 = -i \lambda \left( 1+2\gamma_1 \gamma_2\right)
 e^{i(\omega_1-\omega_2)t}\\
 \dot \gamma_3 = -i \lambda \gamma_1 e^{i(\omega_1-\omega_2)t} .
 \end{eqnarray*}
The time evolution operator $\hat U_I^{(2)}$ satisfies the
equation
\[ i\hbar \partial_t \hat U_I^{(2)} = \left[\hat U_I^{(1)\dagger}\hat V_2 \hat U_I^{(1)}\right]\hat
U_I^{(2)}\equiv \hat H_I^{(2)} \hat U_I^{(2)},
\]
and transforming the interaction we obtain:
\begin{eqnarray}\label{eq:h2} 
  \hat H_I^{(2)}&=& \hbar g_1\left[(1+\gamma_1
    \gamma_2)e^{-\gamma_3-i(\Omega_1-\omega_1)t}\hat a_1^{\dagger}\hat
    \sigma_1^{(-)} + e^{\gamma_3+i(\Omega_1-\omega_1)t}\hat a_1 \hat
    \sigma_1^{(+)}\right]\nonumber \\ &+& \hbar
  g_2\left[e^{\gamma_3-i(\Omega_2-\omega_2)t}\hat a_2^{\dagger}\hat
    \sigma_2^{(-)} +
    (1+\gamma_1\gamma_2)e^{-\gamma_3+i(\Omega_2-\omega_2)t}\hat a_2
    \hat \sigma_2^{(+)}\right] \nonumber \\
  &+& \hbar g_1\left[
    \gamma_2 e^{-\gamma_3+i(\Omega_1-\omega_1)t}\hat a_2\hat
    \sigma_1^{(+)}-\gamma_1 e^{\gamma_3-i(\Omega_1-\omega_1)t}\hat
    a_2^{\dagger}\hat \sigma_1^{(-)}\right] \nonumber \\ &+& \hbar
  g_2\left[\gamma_2 e^{-\gamma_3 -i(\Omega_2-\omega_2)t} \hat
    a_1^{\dagger}\hat \sigma_2^{(-)} + \gamma_1
    e^{\gamma_3+i(\Omega_2-\omega_2)t} \hat a_1 \hat
    \sigma_2^{(+)}\right].
\end{eqnarray}
The first two terms correspond to two generalized Jaynes-Cummings
Hamiltonians, one for each cavity and atom. The other two terms,
which are first order in the coefficients $\gamma_i$, involve the interaction between cavity two and the atom in cavity one, and that between cavity one and the atom in cavity two. It is to be expected that these mixing terms will have a minor importance on the dynamics of the system  and we will assume that it is so.  If it were not the case, we would find out when we compare the numerical results with those obtained within this approximation. Then,  keeping only the terms corresponding to generalized Jaynes-Cummings Hamiltonians, we arrive at the approximate Hamiltonian
\begin{eqnarray}
\tilde H_I^{(2)}&=& \hbar g_1\left[(1+\gamma_1
    \gamma_2)e^{-\gamma_3-i(\Omega_1-\omega_1)t}\hat a_1^{\dagger}\hat
    \sigma_1^{(-)} + e^{\gamma_3+i(\Omega_1-\omega_1)t}\hat a_1 \hat
    \sigma_1^{(+)}\right]\nonumber \\ &+& \hbar
  g_2\left[e^{\gamma_3-i(\Omega_2-\omega_2)t}\hat a_2^{\dagger}\hat
    \sigma_2^{(-)} +
    (1+\gamma_1\gamma_2)e^{-\gamma_3+i(\Omega_2-\omega_2)t}\hat a_2
    \hat \sigma_2^{(+)}\right], \nonumber \\
\end{eqnarray}
whose time evolution operator may be written as $\hat U_I^{(2)} =
\hat U_{JC}^{(1)} \hat U_{JC}^{(2)}$.

Now we introduce the operators \cite{moya, cordero1}
\[ \hat b_i = \frac{1}{\sqrt{\hat M_i}}\hat a_{i} \hat \sigma_{i}^{(+)},\ \ \
\hat b_i^{\dagger} = \hat a_i^{\dagger}\hat
\sigma_{i}^{(-)}\frac{1}{\sqrt{\hat M_i}} \] with $\hat M_i = \hat
n_i + \frac{1}{2}(1+\hat \sigma_i^{(z)})$ the total number of
excitations in a given ladder. These operators acting upon the
basis states yield
\[ \hat b_i|n_i,e_i\rangle = 0, \ \ \ \hat b_i|n_i+1,g_i\rangle =
|n_i, e_i\rangle \]
\[ \hat b_i^{\dagger}|n_i, e_i\rangle = |n_i+1,g_i\rangle, \ \ \ \hat
b_i^{\dagger}|n_i+1,g_i\rangle = 0 \]
\[ \hat M_i|n_i, e_i\rangle =(n_i+1)|n_i,e_i\rangle, \ \ \ \hat M_i
|n_i+1, g_i\rangle = (n_i+1)|n_i+1,g_i\rangle. \] Notice that
\[ \hat M_i|0,g_i\rangle = 0, \ \ \ \hat b_i|0,g_i\rangle =0 \]
and $\hat b_i^2$, $\hat b_i^{\dagger 2}$ acting upon any state of
the basis is zero. From the above expressions we obtain the
commutation relations
\begin{equation}
  [\hat b_i,\hat b_i^{\dagger}] = \hat \sigma_i^{(z)}, \ \ [\hat
    \sigma_i^{(z)},\hat b_i]= 2\hat b_i, \ \ [\hat \sigma_i^{(z)},\hat
    b_i^{\dagger}]= -2\hat b_i^{\dagger}.
  \end{equation}
The interaction Hamiltonian can be written as
\begin{equation}\label{eq:hint2}
  \hat H_I^{(2)} = \hbar g_1\sqrt{\hat M_1}\left[\phi_{1 1}(t)\hat
    b_1^{\dagger}+\phi_{1 2}(t)\hat b_1\right]+\hbar g_2\sqrt{\hat
    M_2}\left[\phi_{2 1}(t)\hat b_2^{\dagger}+\phi_{2 2}(t)\hat
    b_2\right]
  \end{equation}
whose exact time evolution operator is:
\begin{equation}\label{eq:opevol}
  U_I^{(2)} = \exp[\beta_1^{(z)}\hat
    \sigma_1^{(z)}]\exp[\beta_1^{(+)}\hat
    b_1^{\dagger}]\exp[\beta_1^{(-)}\hat b_1]\exp[\beta_2^{(z)}\hat
    \sigma_2^{(z)}] \exp[\beta_2^{(+)}\hat
    b_2^{\dagger}]\exp[\beta_2^{(-)}\hat b_2]
\end{equation}
with complex, time dependent functions $\beta_i^{(z)}$,
$\beta_i^{(\pm)}$ with initial conditions
$\beta_i^{(z)}(t_0)=\beta_i^{(\pm)}(t_0) = 0$. These functions
satisfy a set of coupled ordinary differential equations obtained
after substitution of Eq.~\ref{eq:opevol} in Schr\"odinger's
equation.

Notice that for the state $|0,g_i\rangle$,  $M_i=0$ and $\dot
\beta_i^{(z)} = \dot \beta_i^{(\pm)} = 0$ and the operator $\hat
U_I^{(2)}$ is the identity operator.

The full time evolution operator is then written as a product 
\begin{equation}
  \hat U = \hat U_0 \hat U_I^{(1)} \hat U_{JC}^{(1)} \hat U_{JC}^{(2)}.
\end{equation}

Since the JC Hamiltonian conserves the total number of excitations
in a given cavity then, the basis states are the direct product of
states $|n_i,e_i\rangle$ and $|n_i+1,g_i\rangle$,  $i=1,2$. At the
initial time, the state of the system is written in terms of the
basis states with fixed number of excitations $M_1$ and $M_2$.
When we apply the time evolution operator, the part corresponding
to the coupling between the cavities modifies these numbers but
conserves their sum $M=M_1+M_2$, then the number of excitations in
each cavity is a function of time. This is taken into account when
we construct the interaction Hamiltonian $\hat{H}_I^{(2)}$ given in
Eq.~(\ref{eq:hint2}).

\section{Numerical results}
In order to illustrate the methodology consider an  initial state
given as $|\Psi(0)\rangle = |n_1, e_1\rangle \otimes |n_2+1,
g_2\rangle$,  that is, we assume that cavity one has $n_1$
excitations and the atom is in its excited state and cavity two
has $n_2+1$ excitations and its atom is in its ground state. We
first apply the evolution operator $\hat U_{JC}^{(1)} \hat
U_{JC}^{(2)}$ to the initial state. As a result we obtain
  \begin{equation}
    |\Psi(1)\rangle = \hat U_{JC}^{(1)}|n_1,e_1\rangle \otimes \hat
    U_{JC}^{(2)}|n_2+1,g_2\rangle,
  \end{equation}
  where:
\[ \hat U_{JC}^{(1)}|n_1,e_1\rangle = e^{\beta_1^{(z)}}|n_1,e_1\rangle
+ \beta_1^{(+)} e^{-\beta_1^{(z)}}|n_1+1,g_1\rangle, \] and
\[ \hat U_{JC}^{(2)}|n_2+1,g_2\rangle =
e^{-\beta_2^{(z)}}(1+\beta_2^{(+)}\beta_2^{(-)})|n_2+1, g_2\rangle
+ e^{\beta_2^{(z)}}\beta_2^{(-)} |n_2,e_2\rangle, \] so that the
state is:
\begin{eqnarray}
  |\Psi(1)\rangle &=& C_{n_1,n_2+1}^{e_1,g_2}|n_1,e_1\rangle \otimes
  |n_2+1,g_2\rangle + C_{n_1+1, n_2+1}^{g_1, g_2}|n_1+1,g_1\rangle
  |\otimes |n_2+1,g_2\rangle \nonumber \\ &+& C_{n_1, n_2}^{e_1,e_2}
  ||n_1,e_1\rangle\otimes |n_2,e_2\rangle + C_{n_1+1,
    |n_2}^{g_1,e_2}|n_1+1,g_1\rangle \otimes |n_2,e_2\rangle.
  \end{eqnarray}
After this part of the evolution, the atoms in each cavity are in
states that are combinations of ground and excited states.

Now apply the operator $\hat U_I^{(1)}$ given in
Eq.~\ref{eq:opev1} to the state $|\Psi(1)\rangle$.  We use the
notation $|n_1,e_1\rangle\otimes|n_2+1,g_2\rangle =
|n_1,n_2+1\rangle |e_1,g_2\rangle$ since this operator does not
involve the atomic degrees of freedom and it only applies upon
field states $|n_1,n_2\rangle$.
\begin{eqnarray}
  \hat U_I^{(1)}|n_1,n_2\rangle &=& \sqrt{\frac{n_2!}{n_1!}}
  e^{\gamma_3(n_1-n_2)} \sum_{p=0}^{n_2}
  \frac{\gamma_2^p}{p!}\frac{(n_1+p)!}{(n_2-p)!}\nonumber \\
  &\times& \sum_{k=0}^{n_1+p}
  \frac{\gamma_1^k}{k!}\sqrt{\frac{(n_2-p+k)!}{(n_1+p-k)!}}
  |n_1+p-k,n_2-p+k\rangle
  \end{eqnarray}
notice that the summations are finite. We have one of these terms
for each of the four different atomic states that appear in
$|\Psi(1)\rangle$.

Then we have:
\begin{eqnarray}
  |\Psi(2)\rangle &=& \hat{U}_I^{(1)}|\Psi(1)\rangle \nonumber \\
 &=&C_{n_1,n_2+1}^{e_1,g_2} U_I^{(1)}|n_1,n_2+1\rangle |e_1,g_2\rangle
  + C_{n_1+1, n_2+1}^{g_1,g_2} U_I^{(1)}|n_1+1,n_2+1\rangle
  |g_1,g_2\rangle \nonumber \\ &+& C_{n_1,n_2}^{e_1,e_2}
  U_I^{(1)}|n_1,n_2\rangle |e_1,e_2\rangle +
  C_{n_1+1,n_2}^{g_1,e_2}U_I^{(1)} |n_1+1,n_2\rangle|g_1,e_2\rangle.
  \end{eqnarray}
Finally we apply the operator $\hat U_0$ to the state
$|\Psi(2)\rangle$.

In order to test the validity of our approximations, we considered
an specific example. Take as initial state $|\Psi(0)\rangle =
|0,e_1\rangle \otimes |2,g_2\rangle$, that is, cavity one has no
photons and the atom is in its excited state and cavity two has
two photons and its atom is in its ground state so that we have a
total of three excitations in the system. Due to the fact that the
Jaynes-Cummings part of the interaction conserves the total number
of excitations in each cavity and the coupling between the
cavities conserves the total number of photons, then the number
$M=M_1+M_2$ giving the total number of excitations for the
complete system is constant.

The wave function after application of the full time evolution
operator can be written as:

\begin{equation}
  |\Psi(t)\rangle =
  \sum_{n_1=0}^3\sum_{n_2=0}^3\sum_{s_1=0}^1\sum_{s_2=0}^1
  \phi_{n_1,n_2}^{s_1,s_2}(t)|n_1,s_1\rangle \otimes |n_2,s_2\rangle
\end{equation}
with the condition $n_1+n_2+s_1+s_2=3$.

For the numerical evaluation of the temporal evolution of the
system we took Hamiltonian parameters given in Ref~\cite{solano},
that is: the cavity frequencies $\omega_1/2\pi= 4$GHz,
$\omega_2/2\pi=5$GHz, the atom-cavity couplings $g_1 =
.04\omega_1$, $g_2= .04\omega_2$, the cavity-cavity coupling
$\lambda = 10^{-3}\omega_1$ and for the atomic energies
$\Omega_1=0.999 \omega_1$, $\Omega_2=0.999\omega_2$. Time will be
given in units of the period of cavity one ($T_1=0.25\times
10^{-9}$s).

\begin{figure}[h!]
\begin{center}
\includegraphics[width=\linewidth]{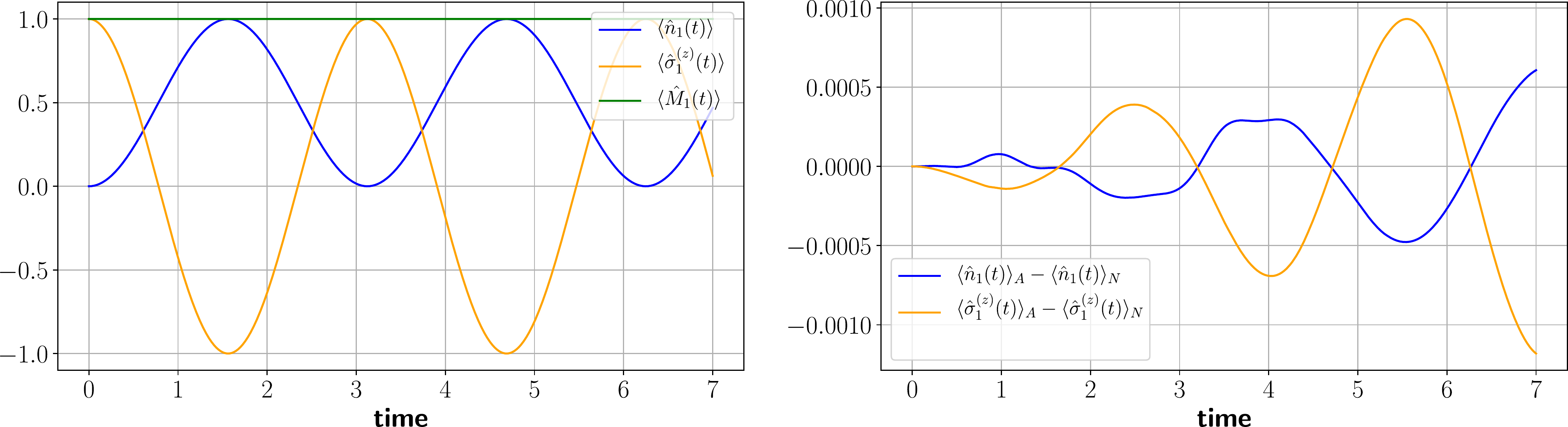}
\caption{Left panel: Average value of the number operator $\langle\hat{n}_1\rangle_A$
in blue and of $\langle\hat{\sigma}_1^{(z)}\rangle_A$ in yellow. The
number of excitations $\langle\hat{M}_1(t)\rangle_A$ is shown in green. Right panel: Difference between the analytic and the numerical calculation, $\langle\hat{n}_1\rangle_A-\langle\hat{n}_1\rangle_N$ in blue and $\langle\hat{\sigma}_1^{(z)}\rangle_A -\langle\hat{\sigma}_1^{(z)}\rangle_N$ in yellow. 
Parameters set $\Omega_1=.999\omega_1$, $\Omega_2=.999\omega_2$,
$\omega_1/2\pi=4$GHz, $\omega_2/2\pi=5$GHz, $g_1=0.04\omega_1$,
$g_2=0.04\omega_2$ and $\lambda= 10^{-3}\omega_1$.}
\label{fig1art}
\end{center}
  \end{figure}

In the left panel of figure \ref{fig1art},  we show the temporal evolution of the
average number of photons  $\langle \hat{n}_1\rangle_A$, the average value
of $\langle \hat{\sigma}_1^{(z)}\rangle_A$ as well as the average value of
the total number of excitations in cavity one $\langle
\hat{M}_1(t)\rangle_A$ calculated with the analytic method. We see that $\langle \hat{M}_1(t)\rangle_A$ remains
constant along the evolution, this means that the cavity-cavity
coupling is too small and we have two almost independent Jaynes-Cummings
Hamiltonians. Since the cavities are almost resonant
$\Omega_1\simeq \omega_1$ the atom-field exchange is complete. The
behavior of cavity two is similar. We see then that with this set
of Hamiltonian parameters there will not be any photon exchange
between the cavities. We also made a purely numerical calculation
using Python \cite{python} and found a very good agreement between
both  calculations as can be seen in the right panel of the figure where the differences are at most of the order of $10^{-3}$.

\begin{figure}[h!]
\begin{center}
\includegraphics[width=\linewidth]{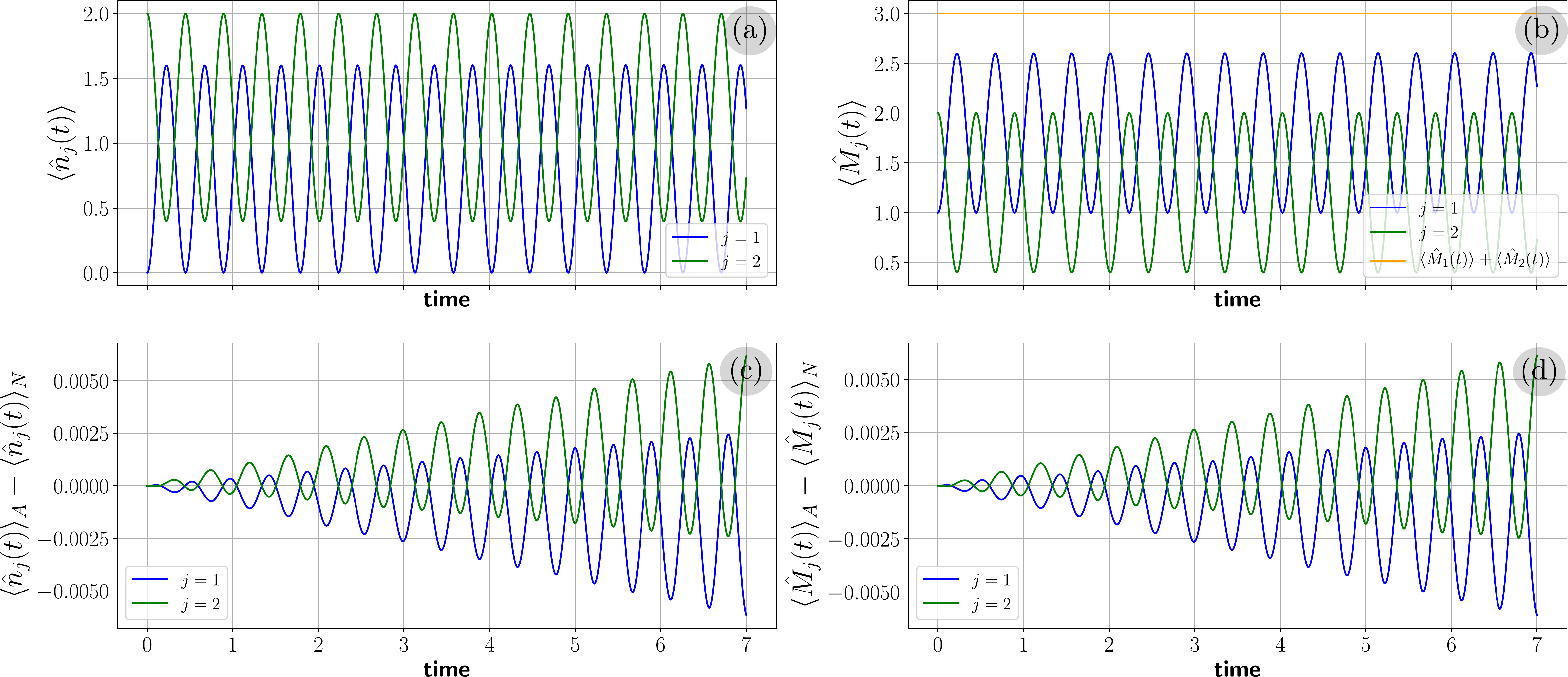}
\caption{First row: (a) Average value of the number operator $\langle\hat{n}_1\rangle_A$
in blue and of $\langle\hat{n}_2\rangle_A$ in green. (b) Average value of the total number of excitations in
cavity  one $\langle\hat{M}_1\rangle_A$ in blue, in cavity two $\langle\hat{M}_2\rangle_A$
  in green and in the full system in yellow. Second row: (c) Difference between the analytic and the numerical calculations for the number operator, (d) Difference between the analytic and the numerical calculations for the excitations.  Parameters set
$\Omega_1=.999\omega_1$, $\Omega_2=.999\omega_2$,
$\omega_1/2\pi=4$GHz, $\omega_2/2\pi=5$GHz, $g_1=0.001\omega_1$,
$g_2=0.001\omega_2$ and $\lambda= .25\omega_1$.} \label{fig3art}
\end{center}
  \end{figure}

In figure \ref{fig3art} (a) we show the temporal evolution of $\langle
\hat n_1(t)\rangle$ (blue) and $\langle \hat n_2(t)\rangle$
(green) calculated with the analytic method for Hamiltonian parameters $\Omega_1=.999\omega_1$,
$\Omega_2=.999\omega_2$, $\omega_1/2\pi=4$GHz,
$\omega_2/2\pi=5$GHz, $g_1=0.001\omega_1$, $g_2=0.001\omega_2$ and
$\lambda= 0.25 \omega_1$.  Here, the atom-field coupling in each
cavity is small and there is almost no exchange between the field
and the atomic excitations in each cavity. On the other hand, the
cavity-cavity coupling is large and there is an important exchange of
photons between the cavities. In figure \ref{fig3art} (b) we show the
evolution of the total number of excitations in each cavity  as
well as the total number of excitations in the system calculated with the analytic method. Since the
cavity-cavity coupling conserves the total number of photons and
the JC Hamiltonians conserve the number of excitations in each
cavity, there is an exchange of excitations between the cavities
keeping the total number of excitations constant.  For this set of
Hamiltonian parameters we also made a purely numerical 
calculation and both calculations yield almost the same results as can be seen in (c) and (d) where we plot the differences between the analytic and the numerical calculations. We can
then be sure that the approximations made along the way are
justified when either the atom-field coupling is large compared with the cavity-cavity coupling (figure one) and also when the atom-field coupling is small 
compared with the cavity-cavity coupling (figure 2).

\begin{figure}[h!]
\begin{center}
\includegraphics[width=\linewidth]{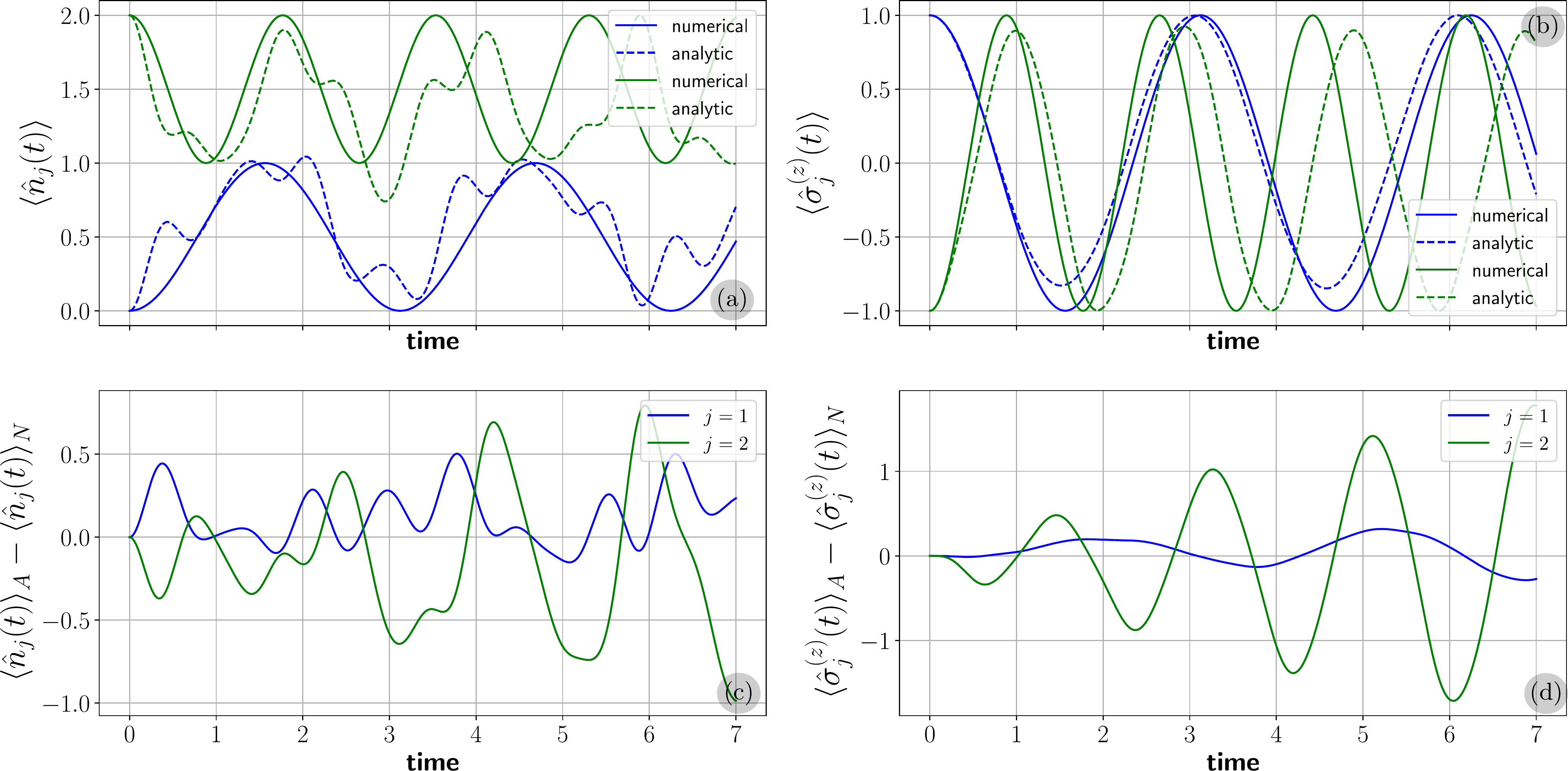}
\caption{First row:Average value of the number operator $\langle
\hat{n}_1\rangle$ in blue and of $\langle\hat{n}_2\rangle$ in green (left panel).
Average value of $\langle\hat{\sigma}_1^{(z)}\rangle$ in blue and
$\langle\hat{\sigma}_2^{(z)}\rangle$ in green (right panel). Second row: Difference between the numerical and the analytic calculations. Cavity one in blue, cavity two in green. Parameters set
$\Omega_1=.999\omega_1$, $\Omega_2=.999\omega_2$,
$\omega_1/2\pi=4$GHz, $\omega_2/2\pi=5$GHz, $g_1=0.04\omega_1$,
$g_2=0.04\omega_2$ and $\lambda= 0.08\omega_1$.} \label{fig5art}
\end{center}
  \end{figure}

Now we consider a case when both coupling parameters are of the
same order of magnitude. Take for instance $g_1=0.04\omega_1$,
$g_2=0.04\omega_2$ and $\lambda = 0.08\omega_1$. We can see the
results in figure \ref{fig5art}. In the first row we plot (left panel) $\langle \hat
n_1\rangle$ and $\langle \hat n_2\rangle$  and (right panel) $\langle
\hat{\sigma}_1^{(z)}\rangle$, $\langle\hat{\sigma}_2^{(z)}\rangle$. Full lines correspond to the analytic calculations and broken lines to the numerical calculation  using Python \cite{python}.  In the second row we plot the difference between the analytic and the numerical calculations.

The behavior of the average number of photons $\langle
\hat{n}_1\rangle$ and $\langle\hat{n}_2\rangle$ differ significantly with the previous results since in this case the coupling between the cavities is large, as well as  the atom-cavity coupling in each one of the cavities.  
In the right panel we see that the averages $\langle\hat{\sigma}_1^{(z)}\rangle$ and $\langle
\hat{\sigma}_2^{(z)}\rangle$ are oscillating functions of time with a nearly constant amplitude. For both atoms there is a shift in the frequency of the oscillations between the analytic and the numerical calculation. For the atom in cavity one, the frequency (analytic) gets larger while for the atom in cavity two it gets smaller and  this means that when $g_i\simeq\lambda$ the approximation made in the interaction Hamiltonian $\hat{H}_I^{(2)}$ in Eq.~\ref{eq:h2} is valid for short times. 
In the second row of the figure we see that the analytic  and the numerical results agree quantitatively for short times and there is only  a qualitative agreement between both  calculations for longer times.

In the following  table,  we resume the behavior of our analytical method compared with the purely numerical calculation stating the values for the Hamiltonian parameters used.
\\
In all the cases we used: $\omega_1/2\pi = 4$ GHz, $\omega_2/2\pi = 5 $GHz, $\Omega_1= 0.999\omega_1$, and $\Omega_2=0.999\omega_2$. 

\begin{center}
\begin{tabular}{|c|c|c|c|c|}
\hline
$g_1$&$g_2$&$\lambda$&Quantit. and Qualit agreement&Qualitative only\\
\hline
$.04\omega_1$&$.04\omega_2$&$10^{-3}\omega_1$&XX& \\
\hline
$.001\omega_1$&$.001\omega_2$&$.25\omega_1$&XX& \\
\hline
$.01\omega_1$&$.01\omega_2$&$.02\omega_1$&XX& \\
\hline
$.4\omega_1$&$.4\omega_2$&$.001\omega_1$&XX& \\
\hline
$.04\omega_1$&$.04\omega_2$&$.08\omega_1$& &XX \\
\hline
\end{tabular}
\end{center}

\section{Conclusions}
In this work we have presented a method to construct an approximate
time evolution operator, written in a product form, for a system composed of two coupled
cavities containing a two level atom each (two coupled Jaynes
Cummings Hamiltonians). The coupling between the cavities conserves the total number of
excitations in the system; it creates an excitation in one cavity
and kills another one in the other cavity. For uncoupled Jaynes
Cummings Hamiltonians the problem is exactly solvable as well as
for two coupled cavities with no atoms in them. However, when
there are atoms present in the cavities and a coupling between
them the problem does not have an exact solution. In order to tackle
the problem we first took as unperturbed Hamiltonian the free
fields and the two two-level atoms, then we transformed the
interaction and obtained a Hamiltonian
that could be written as the sum of two independent JC
Hamiltonians and a coupling between them. We built the time
evolution operator corresponding to the coupling and used it to
transform the rest of the interaction. Finally, we approximated this last  interaction Hamiltonian as
the sum of two independent JC Hamiltonians whose time evolution
operators could  be written {\em exactly} in a product form. 

We want to stress the fact that the approximations are made at the level of the interaction Hamiltonian in order to have 
an approximate Hamiltonian that can be written as a linear combination of operators that form a finite Lie algebra. Then, the
corresponding time evolution operator can be obtained exactly by means of the Wei-Norman theorem.
Once we have an analytic expression for the time evolution operator, we can propagate any initial state and compute whatever property we are interested on,  for example the swapping probability, the Mandel parameter, the Husimi function, the density matrix or the average value of any observable.

In order to test the accuracy of our approximations, we considered
a particular initial state and applied to it the full time
evolution operator. We evaluated the average number of photons 
as well as the average value for the atomic state in each cavity and
confronted the results with those obtained by a purely numerical
calculation solving Scr\"odinger's equation for the full Hamiltonian given by Eq.~\ref{eq:hamiltonian}. The Hamiltonian parameters for the
atom-field coupling and the cavity-cavity coupling were taken from
ref.~\cite{solano} and correspond to actual experimental
possibilities. With these Hamiltonian parameters we found a very
good agreement with the numerical calculation and a negligible swapping
probability. From a comparison with our numerical results we can
say that our analytic method can be applied when the atom-field
coupling parameters $g_i\ll \lambda$ with $\lambda\simeq
10^{-1}\omega_1$, also in the oposite limit when the cavity-cavity
coupling constant is much smaller than the atom-field coupling ($\lambda \ll g_i$ with $g_i\simeq 10^{-1}\omega_1$) and when both sets of parameters are of the same order of magnitude
but with the condition of being much smaller than $\omega_1$ (of
the order of $10^{-3}\omega_1$). When neither of these conditions
apply, our method still gives reasonable results at short times and only qualitative
agreement for longer times. \vspace{1cm}

{\bf Acknowledgement: } A. Paredes acknowledges postdoctoral
support from DGAPA UNAM, and we thank Reyes Garc\'{\i}a for the
maintenance of our computers. We acknowledge partial support from
Direcci\'on General de Asuntos del Personal Acad\'emico,
Universidad Nacional Aut\'onoma de M\'exico (DGAPA UNAM) project
PAPIIT IN111119.


\begin{thebibliography}{40}
\bibitem{Jaynes} E.~ T.~ Jaynes and F.~ W.~ Cummings, Proc. IEEE
    {\bf 51}, 89 (1963).
    \bibitem{Shore} B.~ W.~ Shore and P.~ L.~Knight, J. of
      Mod. Opt.{\bf 40}, 1195 (1993).
      \bibitem{Haroche1} Bertet, P., Auffeves, A., Maioli, P.,
        Osnaghi, S., Meunier, T., Brune, M., Raimond, J.~ M., and
        Haroche, S. Phys. Rev. Lett. {\bf 89}, 200402 (2002).
        \bibitem{Braak} D.~ Braak, Phys. Rev. Lett. {\bf 107}, 100401
          (2011).
           \bibitem{Chen} J.~ Chen, D. Konstantinov and K. Molmer,
            Phys. Rev. A {\bf 99}, 013803 (2019).
          \bibitem{Glauber} R.J. Glauber, Phys. Rev. {\bf 131}, 2766-2788 (1963).
          \bibitem{Kurizki} B. Sherman and G. Kurizki, Phys. Rev. A
            {\bf45}, R7674 (1992).
          \bibitem{Schleich}K. Vogel, V.~M.~ Akulin, and W.~
            P.~Schleich, Phys. Rev. Lett. {\bf 71}, 1816 (1993).
          \bibitem{Dantsker}H. Moya-Cessa, P. L. Knight, and
            A. Rosenhouse-Dantsker, Phys. Rev. A {\bf 50}, 1814-1821 (1994).
          \bibitem{Kuklinski}J. R. Kuklinski, J. L. Madajczyk,
            Phys. Rev. A {\bf 37}, 3175-3178 (1988).
          \bibitem{Vogel1} R. L. Matos Filho and W. Vogel,
            Phys. Rev. Lett. {\bf 76}, 608 (1996).
          \bibitem{Vogel2} R. L. Matos Filho and W. Vogel,
            Phys. Rev. A {\bf 54}, 4560 (1996).
           \bibitem{cordero1} S.~Cordero and J.~R\'ecamier, J. Phys. A:
              Math. Theor. {\bf 45}, 385303 (2012).
              \bibitem{cordero2} S.~ Cordero and J.~ R\'ecamier,
                J. Phys. B: At. Mol. Opt. Phys. {\bf 44}, 135502
                (2011).
                \bibitem{santos1} O. de los Santos-S\'anchez, and
                  J. R\'ecamier, J. Phys. B: At. Mol. Opt. Phys. {\bf 45}, 015502
                  (2012).
          \bibitem{Fili} P. Filipowicz, J. Javanainen, P. Meystre,
            Phys. Rev. A {\bf 34}, 3086 (1996).
\bibitem{Power} H. Moya-Cessa, V. Buzek and P.L. Knight, Opt. Commun. {\bf 85}, 267-274 (1991).
            \bibitem{moore} G. T. Moore, J. Math. Phys. {\bf 11}, 2679-2691 (1970).
            \bibitem{nation} P. D. Nation, J. R. Johansson, M. P. Blencowe and F. Nori, Rev. Mod. Phys. {\bf 84}, 1-24 (2012).
            \bibitem{dodonov} A. V. Dodonov and V. V. Dodonov, Phys. Rev. A {\bf 85}, 055805 (2012).
            \bibitem{yan} Yan-Ling Li, Mao-Fa Fang, Xing Xiao, Ke Zeng and Chao Wu, J. Phys. B:At. Mol. Phys. {\bf 43}, 085501 (2010).
            \bibitem{doherty} H. Mabuchi and A. C. Doherty, Science {\bf 298}, 1372 (2002).
            \bibitem{hassel} P. L\"ahteenm\"aki, G. S. Paraoanu, J. Hassel and P. J. Hakonen, Proc. Natl. Acad. Sci. USA {\bf 110}, 4234-4238 (2013).
            \bibitem{nori} C. M. Wilson, G. Johansson, A. Pourkabirian, M. Simoen, J. R. Johansson, F. Duty, T. Nori and P. Delsing, Nature {\bf 479}, 376-379 (2011). 
          \bibitem{solano} S. Felicetti, M. ~ Sanz, L.~ Lamata, G.~ Romero, G.~
            Johansson, P.~ Delsing and E.~ Solano,  Phys. Rev. Lett. {\bf 113}, 093602
            (2014).
            \bibitem{Hanoura} S. A. Hanoura, M. M. A. Ahmed,
              E. M. Khalil, and A. S. F. Obada, Fortschr. Phys. {\bf
                67}, 1800101 (2019).
\bibitem{James} D. F. V. James and J. Jerke, Can. J. Phys. {\bf  85}, 625 (2007).
            \bibitem{wei-norman}J.~ Wei and E.~ Norman, Proc. Am. Math. Soc. {\bf 15},
              327-334 (1964).
            \bibitem{moya} B.~M.~Rodr\'{\i}guez-Lara, H.~M.~Moya-Cessa, J. Phys. A: Math. Theor. {\bf 46}, 095301 (2013).
             \bibitem{python} J.~R.~Johansson, P.~D.~Nation, and F.~Nori,
              "QuTiP2: A Python framework for the dynamics of open quantum
              systems", Comp. Phys. Comm. {\bf 184}, 1234 (2013).
\end{thebibliography}
\end{document}